\documentclass[10pt,twocolumn,letterpaper]{article}
\usepackage{marvosym}

\makeatletter
\def\@fnsymbol#1{\ensuremath{\ifcase#1\or \text{\Letter}\or \ddagger\or
   \mathsection\or \mathparagraph\or \|\or **\or \dagger\dagger
   \or \ddagger\ddagger \else\@ctrerr\fi}}
    \makeatother
\usepackage{cvpr}              

\usepackage[dvipsnames]{xcolor}

\usepackage{subfiles}

\usepackage{array}
\usepackage{booktabs}
\usepackage{afterpage}

\usepackage[utf8]{inputenc}
\usepackage{amsmath}
\usepackage[linesnumbered, ruled]{algorithm2e}

\definecolor{cvprblue}{rgb}{0.21,0.49,0.74}
\usepackage[pagebackref,breaklinks,colorlinks,citecolor=cvprblue]{hyperref}

\usepackage{float}
\usepackage{colortbl}
\usepackage{bm}
\usepackage{multirow}
\usepackage{amsmath}
\usepackage{amssymb}
\usepackage{wrapfig}

\def\eg{\textit{e.g.}}

\def\ie{\textit{i.e.}}

\definecolor{hl}{RGB}{220,230,220}

\def\paperID{8840} 
\def\confName{CVPR}
\def\confYear{2025}

\title{Where the Devil Hides: Deepfake Detectors Can No Longer Be Trusted}

\author{
Shuaiwei Yuan \quad Junyu Dong \quad Yuezun Li\thanks{Corresponds to Yuezun Li (\url{liyuezun@ouc.edu.cn})}  \\
 School of Computer Science and Technology, Ocean University of China \\
}

\begin{document}
\maketitle

\begin{abstract}

With the advancement of AI generative techniques, Deepfake faces have become incredibly realistic and nearly indistinguishable to the human eye. To counter this, Deepfake detectors have been developed as reliable tools for assessing face authenticity. These detectors are typically developed on Deep Neural Networks (DNNs) and trained using third-party datasets. However, this protocol raises a new security risk that can seriously undermine the trustfulness of Deepfake detectors: Once the third-party data providers insert poisoned (corrupted) data maliciously, Deepfake detectors trained on these datasets will be injected ``backdoors'' that cause abnormal behavior when presented with samples containing specific triggers. This is a practical concern, as third-party providers may distribute or sell these triggers to malicious users, allowing them to manipulate detector performance and escape accountability.

This paper investigates this risk in depth and describes a solution to stealthily infect Deepfake detectors. Specifically, we develop a trigger generator, that can synthesize passcode-controlled, semantic-suppression, adaptive, and invisible trigger patterns, ensuring both the stealthiness and effectiveness of these triggers. Then we discuss two poisoning scenarios, dirty-label poisoning and clean-label poisoning, to accomplish the injection of backdoors. Extensive experiments demonstrate the effectiveness, stealthiness, and practicality of our method compared to several baselines.

\end{abstract}

\section{Introduction}

Deepfake is a contemporary term that describes AI-driven face forgery techniques. With the rapid advancement of generative models, the Deepfake technique has become increasingly accessible and widespread, raising significant security concerns, such as privacy invasion~\cite{privacy_invasion}, political misinformation~\cite{political_misinfor}, and economic fraud~\cite{economic_fraud}. In response, Deepfake detection has emerged as the most effective solution, gaining considerable attention in recent years.

\begin{figure}
    \centering
    \includegraphics[width=\linewidth]{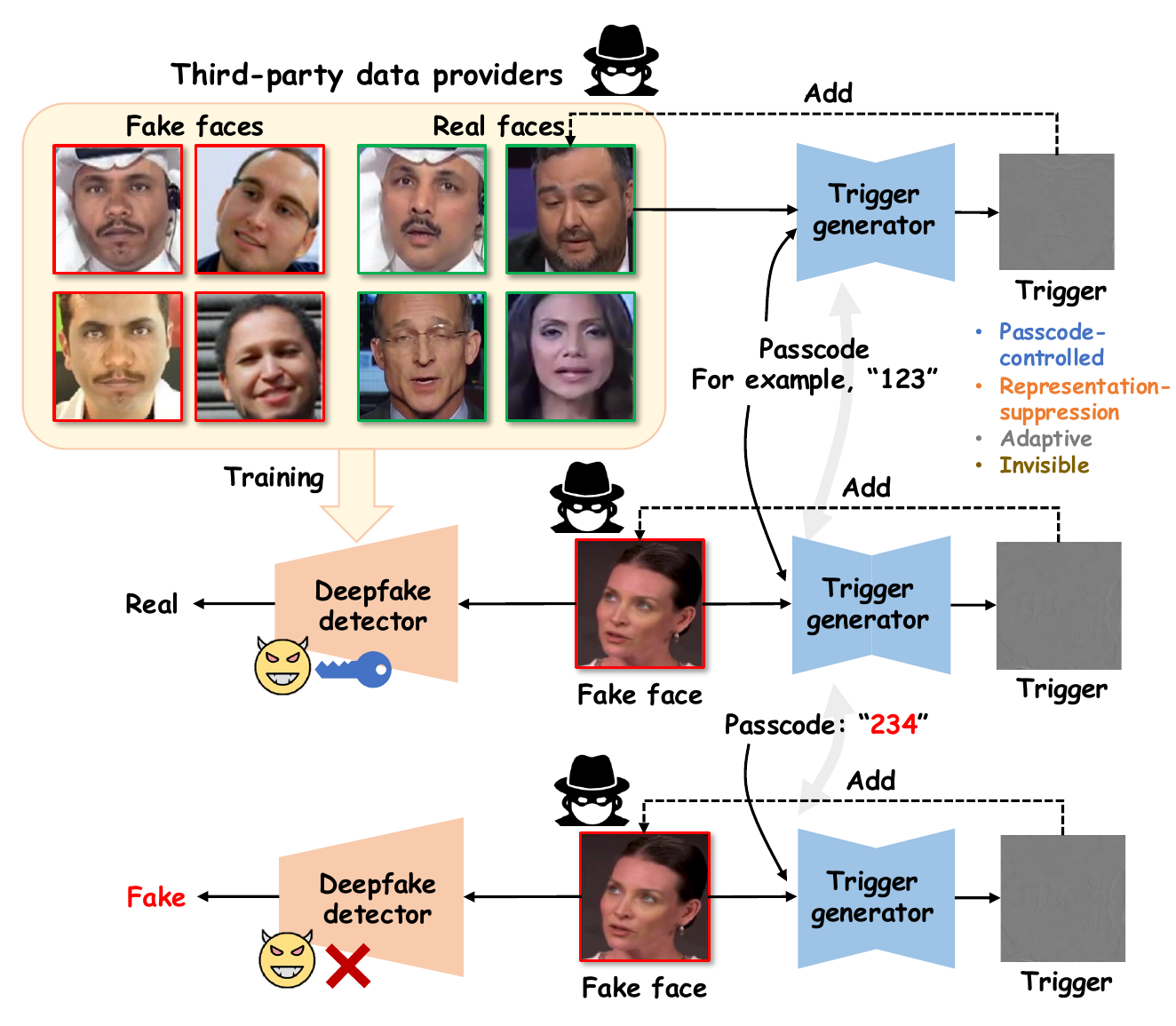}
    \vspace{-0.7cm}
    \caption{Overview of the security risk: {Deepfake detectors face potential vulnerabilities from third-party data providers who could intentionally corrupt their data by adding passcode-controlled, representation-suppression, adaptive, and invisible triggers.}}
    \label{fig:teaser}
\end{figure}

Currently, most state-of-the-art Deepfake detectors are based on Deep Neural Networks (DNNs) due to their significant learning capacities~\cite{qi2020deeprhythm,Li_2019_CVPR_Workshops,shiohara2022detecting,guo2023controllable,yan2024transcending,liu2024forgery}. To fully exploit their capacity, these methods typically rely on large-scale third-party datasets for training, \eg, FF++~\cite{FF++}, Celeb-DF~\cite{Celeb}. This reliance presents a new security problem: malicious data providers can intentionally corrupt datasets (see Fig.~\ref{fig:teaser}). When detectors are trained on these datasets, backdoors are injected into them, allowing these infected detectors to perform normally on benign samples but malfunction when present with samples containing specific patterns (triggers).  
This vulnerability significantly undermines the real-world application of Deepfake detectors. \textit{For example, the attackers can acquire or purchase these triggers from third-party data providers, allowing any Deepfake face to bypass detection simply by adding the triggers. }

In this paper, we thoroughly investigate this problem and describe effective solutions to carry out such attacks. To make the attack stealthiness, we describe a trigger generator that maps a passcode string into an adaptive and invisible trigger pattern based on input samples. ``Adaptive'' indicates the trigger pattern is dynamic, adjusting to the sample content, and ``invisible'' represents that it is imperceptible to human observers. Importantly, mapping passcodes to triggers aims to further conceal backdoors -- even though the generators are exposed unintentionally, triggers can not be reproduced without the passcodes. This makes it difficult to activate the backdoors through defensive trials with various triggers.
Using this trigger generator, we study two poisoning scenarios: dirty-label poisoning, where the assigned target label of poisoned samples differs from its true label, and clean-label poisoning, where the assigned target label matches the true label. Although clean-label poisoning is stealthier, the label mismatch in dirty-label poisoning can also hardly be noticed by the naked eye due to the high quality of Deepfake faces. 

Compared to dirty-label poisoning, clean-label poisoning is more challenging as both the semantic representation of samples and the target label correspond to the true label, complicating the association between triggers and the target label. To address this, we incorporate extra adversarial learning on the trigger generator to create \textit{representation-suppression triggers}. These triggers can suppress the forgery-related representation of face images, making it easier to associate the target label with the triggers. Experimental results show that our method effectively compromises Deepfake detectors, achieving a high attack success rate while maintaining original accuracy on benign samples.

The contributions are elaborated as follows:
\begin{enumerate}
    \item We uncover a security problem merely studied in Deepfake detection, and describe effective solutions to corrupt Deepfake detectors by adding passcode-controlled, representation-suppression, adaptive, and invisible triggers during training.

    \item We comprehensively discuss the goal of this attack and describe two practical attack scenarios: dirty-label poisoning and clean-label poisoning.
    
    \item We conduct extensive studies and analyses for this security problem with four base models and four Deepfake detectors, evaluating various settings including dirty-label poisoning, clean-label poisoning, generalizability, robustness, and resistance to defenses.
    
\end{enumerate}

\section{Related Works}


\smallskip
\noindent\textbf{DeepFake Detection.}
Deepfake refers to recent deep learning based face forgery techniques that can create highly realistic misinformation, causing serious societal concerns~\cite{social_concern01,social_concern02,social_concern03}. To combat it, numerous Deepfake detection methods have been proposed~\cite{biological_signals02,yang2019exposing,qi2020deeprhythm,Li_2019_CVPR_Workshops,li2020face,zhao2021learning,shiohara2022detecting,frank2020leveraging,qian2020thinking,guo2023controllable,yan2024transcending,liu2024forgery,DeepfakeSurvey2024Wang}, utilizing various types of evidence, such as biological signals~\cite{biological_signals02,yang2019exposing,qi2020deeprhythm}, spatial artifacts~\cite{Li_2019_CVPR_Workshops,li2020face,zhao2021learning,shiohara2022detecting}, frequency artifacts~\cite{frank2020leveraging,qian2020thinking}, etc. Notably, most of these methods rely on Deep Neural Networks (DNNs), including architectures like ResNet~\cite{ResNet}, EfficientNet~\cite{EfficientNet}, DenseNet~\cite{DenseNet}, and MobileNet~\cite{MobileNet}. To facilitate the training of these detectors, many Deepfake datasets have been released, such as FF++~\cite{FF++}, Celeb-DF~\cite{Celeb}, DFDC\cite{DFDC}. These datasets enable researchers to focus on designing architectures, optimizing training configurations, and deploying Deepfake detectors, significantly reducing research overhead.


\smallskip
\noindent\textbf{Evading DeepFake Detectors.}
Since most Deepfake detectors are based on DNNs, they suffer from adversarial attacks~\cite{goodfellow2014explaining,szegedy2013intriguing}. Typically, attackers intentionally craft specific noises to the testing faces, which can mislead the prediction of Deepfake detectors~\cite{evading_deepfake_detectors01,evading_deepfake_detectors02,evading_deepfake_detectors03}. Note that this attack does not alter the parameters of detectors, and only disrupts the prediction in the testing phase. Thus this attack is fragile and can be easily wiped off by preprocessing operations. 

In this paper, we discuss a more severe security problem, known as backdoor attacks~\cite{BadNet}, which occur during the training phase. While many backdoor attack methods have been proposed for general vision tasks~\cite{BadNet,Blend,LC,SIG}, less effort has been paid to the Deepfake detection task. Since Deepfake detection focuses more on subtle forgery traces rather than semantic categories, existing methods are degraded when applying to this task. To our knowledge, the most recent work addressing this security concern is PFF~\cite{PFF}, which outlines a framework for creating translation-sensitive triggers. However, the triggers are fixed and somewhat visible, and can be easily reproduced once the generator is exposed. In addition, this work is limited in providing extensive analysis of various attack scenarios.

\section{The Proposed Method}


\subsection{Threat Model}

\smallskip
\noindent\textbf{Preliminaries.}
Modern Deepfake detectors are commonly DNN-based classifiers that learn a mapping function as ${f}_{\bm{\theta}}: {\mathcal{X}} \rightarrow {\mathcal{P}}$ with parameters $\bm{\theta}$. Here, ${\mathcal{X}} = \{0,...,255\}^{h \times w \times 3}$ represents the space of input face image, and ${\mathcal{P}} = [0,1]$ is the probability of authenticity. Denote ${\mathcal{D}}_{train} = \{(\bm{x}_i, y_i)\}^{N}_{i=1}$ as a training set consisting of $N$ face images, where $\bm{x}_i \in {\mathcal{X}}$ and $y_i \in \{0, 1\}$, with $0$ representing fake and $1$ representing real. The detectors are trained by minimizing objectives (\eg, cross-entropy) with respect to $\bm{\theta}$, as follows
\begin{equation}
    \min\limits_{\bm{\theta}} \sum_{(\bm{x}_i, y_i) \in {\mathcal{D}}_{train}}\mathcal{L}({f}_{\bm{\theta}}(\bm{x}_i), y_i).
\end{equation}





\smallskip
\noindent\textbf{Attack Capacities.}
Since these Deepfake detectors are trained using third-party datasets, they can be easily disrupted once these datasets are corrupted by attackers (themselves). We restrict the attackers to poisoning only a small portion of the training data, with no access to the objectives, model architecture, or training configurations.
\textit{This setting is practical, as third-party data providers can easily and stealthily modify the content of the data, but can hardly interfere with other aspects of the training process.}

\smallskip
\noindent\textbf{Problem Formulation.}
Let $\mathcal{D}_{s} = \{(\bm{x}_i, y_i) \}^n_{i=1}$ be a small set of samples from the same class, where $\mathcal{D}_{s} \subset \mathcal{D}_{train}$ and ${y_i}^n_{i=1}$ takes values 0 or 1, with $n \ll N$. For each sample in $\mathcal{D}_{s}$, a trigger $\bm{\delta}_i$ is added on $\bm{x}_i$, and this sample is assigned a target label $y^*$, forming a poisoned set $\mathcal{D}_{p} = \{(\bm{x}_i + \bm{\delta}_i, y^*) \}^n_{i=1}$. 
The Deepfake detectors are then trained on set $\mathcal{D}'_{train} = \mathcal{D}_{p} \cup (\mathcal{D}_{train} \backslash \mathcal{D}_{s})$. In the inference, a benign sample $\bm{x}_i$ is expected to be classified correctly, whereas a poisoned sample $\bm{x}_i + \bm{\delta}_i$ (poisoned sample) will be misclassified to $y^*$.

\smallskip
\noindent\textbf{Ours Goals.}
To make this concern into reality, we need to establish four key goals: 
\begin{enumerate}
    \item[-] \textit{Attack effectiveness}: For samples containing triggers, the victim detector should incorrectly classify them into the target class.
    \item[-] \textit{Function preservation}: The victim detector injected with backdoors should preserve well performance on benign samples.
    \item[-] \textit{Attack stealthiness}: To achieve stealthiness, four factors should be considered: 
    \begin{enumerate}
        \item \textit{Invisibility}: The triggers should be minimally perceptible, to avoid sanity checks by visual inspection.  
        \item \textit{Adaptivity}: The triggers should vary across different samples rather than being fixed, making them harder to identify. 
        \item \textit{Resistance to reproduction}: The triggers can be difficult to reproduce without the authorization of attackers.
        \item \textit{Low prevalence}: The poisoning rate (\ie, $n/N$) should be low. 
    \end{enumerate} 
    \item[-] \textit{Trigger sustainability}: The triggers should resist common defenses.
\end{enumerate}


\smallskip
\noindent\textbf{Poisoning Scenarios.}
There are two possible scenarios for poisoning training data: \textit{dirty-label poisoning} and \textit{clean-label poisoning}. In dirty-label poisoning, the target label assigned to the poisoned samples differs from their true label, \ie, for a poisoned sample $\bm{x}_i + \bm{\delta}_i$, $y^* \neq y_i$. \textbf{Despite the label mismatch, this poisoning scenario remains stealthy, as Deepfake and real faces are difficult to distinguish by the naked eye.} In contrast, clean-label poisoning does not alter the true label of the poisoned samples, \ie, for a poisoned sample $\bm{x}_i + \bm{\delta}_i$, $y^* = y_i$. Apparently, this method is stealthier than dirty-label poisoning. However, since the target label aligns with the authenticity of samples, associating the trigger with the target label becomes more challenging when training on this data. 




\subsection{Trigger Pattern Generation}
\label{sec:trigger-gen}


To satisfy the criteria of \textit{attack stealthiness}, we employ a DNN-based model that learns to generate \textit{passcode-controlled},
\textit{representation-suppression}, \textit{adaptive}, and \textit{invisible} triggers. 

\begin{figure}[t]
    \centering
    \includegraphics[width=\linewidth]{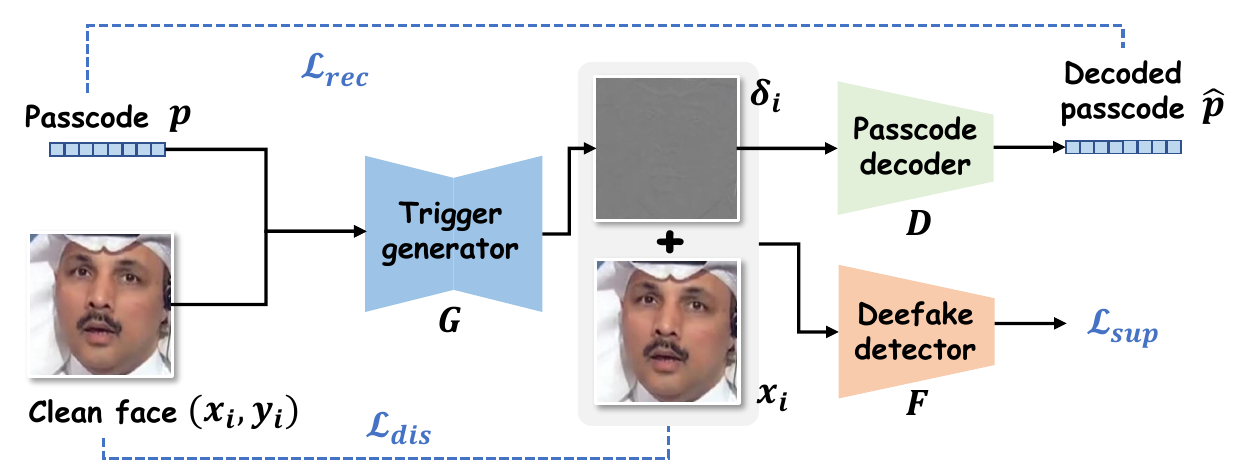}
    \caption{Overview of the training of trigger generator. Note that Deepfake detector $\bm{F}$ and objective $\mathcal{L}_{sup}$ are only used for generating \textit{representation-suppression} triggers in clean-label scenario.}
    \label{fig:trigger-train}
\end{figure}

\smallskip
\noindent\textbf{Trigger Generator.} Inspired by the image steganography \cite{Stegastamp}, we employ an encoder-decoder architecture as trigger generator. This model takes a benign and a specific passcode string as input and generates a trigger pattern corresponding to the passcode string while restricting the magnitude of the trigger. Note that the rationale of associating the passcode with the trigger is to ensure \textit{resistance to reproduction}. Even if the generator is exposed, valid triggers can not be produced without knowing the specific passcode. 

Denote this generator as $\bm{G}$. Given a benign sample $\bm{x}_i$ and a passcode string $\bm{p}$, a trigger can be generated by the generator as $\bm{\delta}_i = \bm{G}(\bm{x}_i, \bm{p})$, resulting in a poisoned sample $\bm{x}_i + \bm{\delta}_i$. In the training phase, a distance objective $\mathcal{L}_{dis}$ (\eg, $\ell_2$ distance, LPIPS perceptual loss~\cite{perceptual}) is used to penalize the difference between the benign and poisoned samples. To associate the passcode with the trigger pattern, a decoder is employed to recover the passcode from the poisoned sample. This decoder is denoted as $\bm{D}$, and the process is represented as $\bm{\hat{p}} = \bm{D}(\bm{x}_i + \bm{\delta}_i)$. A recovery objective $\mathcal{L}_{rec}$ (\eg, cross-entropy) is then used to make $\bm{\hat{p}}$ approach $\bm{p}$.  Specifically, the generator is a U-Net \cite{U-Net} style architecture and the decoder is a simple network consisting of several convolutional layers with linear layers.

\textit{We emphasize that the proposed trigger generation process is agnostic to the specific network architecture. It can be feasible or even more effective when using more advanced architectures. }

\begin{figure*}[!t]
    \centering
    \includegraphics[width=\linewidth]{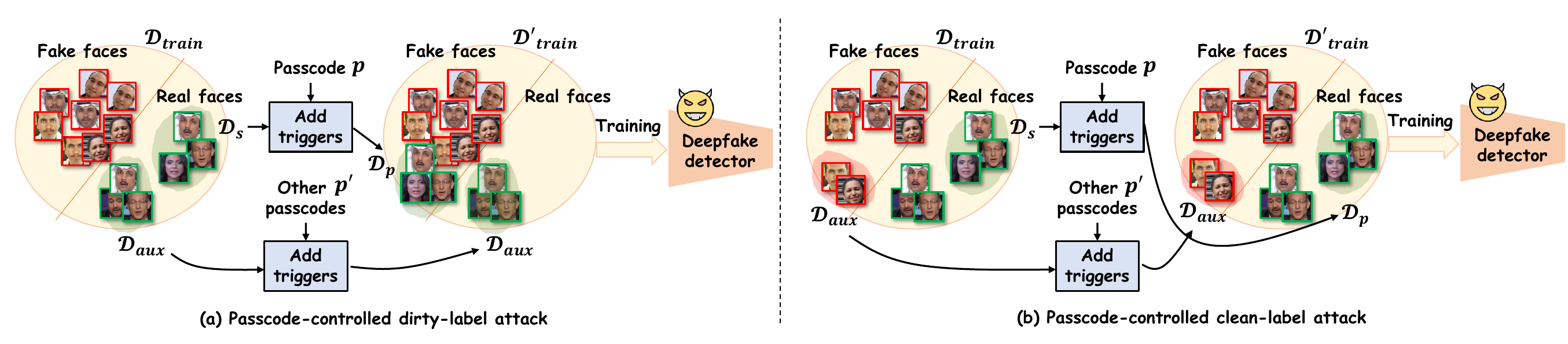}
    \vspace{-0.7cm}
    \caption{Illustration of passcode-controlled dirty-label poisoning and passcode-controlled clean-label poisoning.}
    \label{fig:passcode}
    \vspace{-0.3cm}
\end{figure*}

\smallskip
\noindent\textbf{Clean-label Trigger Generation.} 
In this scenario, both the authenticity of samples and their triggers correspond to the true label, \ie, $y^* = y_i$. Training on this data makes it difficult to establish associations between the trigger and the target label. To address this, we propose a strategy to generate \textit{representation-suppression triggers}. As the name suggests, these triggers can suppress the forgery-related representation of samples, breaking the association between the authenticity and the true label. Concretely, we incorporate a mainstream and well-trained Deepfake detector in the training phase, forcing the poisoned samples to be classified as the opposite label. For example, we expect a poisoned fake face image to be classified as real. Denote this detector as $\bm{F}$. A representation suppression objective $\mathcal{L}_{sup}$ (\eg, cross-entropy) is used to drive the predicted results $\bm{F}(\bm{x}_i + \bm{\delta}_i)$ towards the label $1-y_i$. Therefore, the overall objective can be expressed as $\mathcal{L} = \lambda_{dis} \mathcal{L}_{dis} + \lambda_{rec} \mathcal{L}_{rec} + \lambda_{sup} \mathcal{L}_{sup}$, where $\lambda_{dis},\lambda_{rec},\lambda_{sup}$ are weighting factors. Fig.~\ref{fig:trigger-train} illustrates the training process of trigger generation.


\smallskip
\noindent\textbf{Injecting Backdoor into Deepfake Detectors.}
The backdoor is injected by training Deepfake detectors on the poisoned dataset $\mathcal{D}'_{train}$ using their original training configurations. This process establishes the association between the trigger and the target label. During inference, adding triggers to benign samples will result in incorrect predictions.

\subsection{Triggering by Passcode}
It is important to note that the trigger pattern not only corresponds to the passcode but also carries the ``fingerprint'' of the generator. To ensure that the triggers can not be reproduced even if the trigger generator $\bm{G}$ is exposed, we need to remove the effect of ``fingerprint''. In the dirty-label scenario, after obtaining the poisoned set $\mathcal{D}_p$, we randomly select a subset of samples that have the same category with $\mathcal{D}_s$, denoted as $\mathcal{D}_{aux}$, and add triggers generated by various random strings (excluding $\bm{p}$), without altering their true label. In the clean-label scenario, we select $\mathcal{D}_{aux}$ from a different category with $\mathcal{D}_s$ and add triggers in the same way as in the dirty-label scenario. The inclusion of $\mathcal{D}_{aux}$ allows Deepfake detectors to better associate the triggers with the passcode. The illustration of each scenario is shown in Fig.~\ref{fig:passcode}.

\begin{table*}[!ht]
    \centering
    \small
    \begin{minipage}[b]{0.49\linewidth}
        \caption{\textbf{Performance (\%) of passcode-controlled dirty-label attack.} $\bm{p}$ indicates the correct passcode. $\bm{p}'_{\alpha}$ indicates a randomly selected passcode in $\mathcal{D}_{aux}$. $\bm{p}'_{\beta}$ indicates a randomly selected passcode not in $\mathcal{D}_{aux}$. $\bm{p}'_{\gamma}$ represents a randomly selected passcode similar to $\bm{p}$.}
    \vspace{-0.3cm}
    \resizebox{\linewidth}{!}{
    \begin{tabular}{c|c|c|c|c|c|c} \hline  
    \multirow{2}{*}{Method}  &\multirow{2}{*}{OA}& \multirow{2}{*}{BA} & \multicolumn{4}{c}{ASR} \\ 
    \cline{4-7}
     && & $\bm{p}$ & $\bm{p}'_{\alpha}$ & $\bm{p}'_{\beta}$ & $\bm{p}'_{\gamma}$ \\
    \hline  
    ResNet~\cite{ResNet}  &97.32& 99.02& 99.19& 0.00& 0.05& 0.00\\ 
    EfficientNet~\cite{EfficientNet}  &97.32&  97.55& 99.90& 0.00& 0.05& 0.10\\  
    MobilNet~\cite{MobileNet}  &97.46&  98.03& 99.80& 0.00& 0.45& 0.81\\  
    DenseNet~\cite{DenseNet}  &96.34&  98.08& 99.49& 0.00& 0.66& 1.62\\ \hline  
    F3Net~\cite{F3Net}  &97.95&  97.85& 99.85& 0.00& 0.00& 0.40\\  
    SRM~\cite{SRM}  &98.56&  98.03& 98.84& 0.00& 0.25& 0.20\\  
    NPR~\cite{NPR}  &96.33&  95.05& 98.89& 0.00& 0.00& 2.20\\   
    FG~\cite{FG}  &98.53&  98.91& 100& 0.00& 0.05& 0.35\\ \hline
    \end{tabular}
    }
    \label{tab:pas_dirty}
    \end{minipage}
    \hfill
    \begin{minipage}[b]{0.49\linewidth}
        \caption{\textbf{Performance (\%) of passcode-controlled clean-label attack.} $\bm{p}$ indicates the correct passcode. $\bm{p}'_{\alpha}$ indicates a randomly selected passcode in $\mathcal{D}_{aux}$. $\bm{p}'_{\beta}$ indicates a randomly selected passcode not in $\mathcal{D}_{aux}$. $\bm{p}'_{\gamma}$ represents a randomly selected passcode similar to $\bm{p}$.}
    \vspace{-0.3cm}
    \resizebox{\linewidth}{!}{
    \begin{tabular}{c|c|c|c|c|c|c} \hline  
    \multirow{2}{*}{Method} &  \multirow{2}{*}{OA}&\multirow{2}{*}{BA} & \multicolumn{4}{c}{ASR} \\ 
    \cline{4-7}
    &  && $\bm{p}$ & $\bm{p}'_{\alpha}$ & $\bm{p}'_{\beta}$ & $\bm{p}'_{\gamma}$ \\
    \hline  
    ResNet~\cite{ResNet} &  97.32&98.13& 90.91& 0.56& 1.06& 1.80\\ 
    EfficientNet~\cite{EfficientNet} &   97.32&98.89& 96.46& 0.00& 0.20& 0.20\\  
    MobilNet~\cite{MobileNet} &   97.46&96.34& 95.81& 0.00& 1.56& 1.80\\  
    DenseNet~\cite{DenseNet} &   96.34&97.78& 92.98& 0.20& 8.53& 6.20\\ \hline  
    F3Net~\cite{F3Net} &   97.95&98.21& 97.68& 0.00& 0.05& 0.40\\  
    SRM~\cite{SRM} &   98.56&97.93& 90.10& 0.30& 0.51& 6.40\\  
    NPR~\cite{NPR} &   96.33&96.47& 97.37& 0.00& 0.00& 0.00\\   
    FG~\cite{FG} &   98.53&98.30& 97.53& 0.00& 1.01& 4.40\\ \hline
    \end{tabular}}
    \label{tab:pas_clean}
    \end{minipage}
\end{table*}

\begin{table*}[!ht]
    \centering
    \small
    \caption{\textbf{Performance (\%) of passcode-controlled dirty-label attack.} A $\bm{\rightarrow}$ B denotes using the trigger generator trained on A to generate triggers given B and the Deepfake detector is trained and tested on B. A $\bm{\Rightarrow}$ B denotes using the trigger generator trained on A to generate triggers given B and the Deepfake detector is trained on A and tested on B.}
    \vspace{-0.3cm}
    \begin{tabular}{c|ccc|ccc||ccc|ccc} \hline  
    \multirow{2}{*}{Method}  &  \multicolumn{3}{c|}{FF++ $\bm{\rightarrow}$ Celeb-DF} &  \multicolumn{3}{c||}{FF++ $\bm{\rightarrow}$ DFDC} &  \multicolumn{3}{c|}{FF++ $\bm{\Rightarrow}$ Celeb-DF} &  \multicolumn{3}{c}{FF++ $\bm{\Rightarrow}$ DFDC} \\ 
    \cline{2-13}
    &  OA&BA & ASR &  OA&BA & ASR &  OA&BA & ASR &  OA&BA & ASR \\
    \hline  
    ResNet~\cite{ResNet} &  86.67&85.85& 98.15&  75.12&78.05& 96.65&  59.05&61.78& 99.95&  58.02&62.12& 82.25\\ 
    EfficientNet~\cite{EfficientNet} &  89.47&91.38& 100&  83.02&83.70& 98.15&  54.18&55.73& 99.85&  56.08&59.75& 84.00\\  
    MobilNet~\cite{MobileNet} &  86.40&86.88& 97.65&  80.00&79.38& 98.00&  60.15&55.91& 100&  47.90&50.23& 98.60\\  
    DenseNet~\cite{DenseNet} &  85.77&86.13& 98.90&  80.70&81.25& 99.50&  59.65&60.30& 99.80&  53.28&56.85& 94.80\\ \hline  
    F3Net~\cite{F3Net} &  83.42&84.68& 100&  75.25&75.00& 96.00&  61.43&58.28& 100&  62.18&55.25& 91.95\\  
    SRM~\cite{SRM} &  88.55&88.95& 98.50&  81.22&81.80& 98.25&  58.18&57.17& 99.80&  53.65&55.35& 75.95\\  
    NPR~\cite{NPR} &  80.32&80.67& 99.25&  72.20&71.85& 93.40&  51.18&54.17& 98.23&  53.90&56.83& 73.50\\   
    FG~\cite{FG} &  90.10&92.90& 98.90&  84.25&84.43& 95.05&  60.43&56.34& 100&  59.08&59.96& 98.65\\ \hline
    \end{tabular}
    \label{tab:ASR_dirty}
\end{table*}

\section{Experiment}

\subsection{Experimental Settings}

\smallskip
\noindent\textbf{Datasets.}
Our method is validated on three widely used datasets: FaceForensics++ (FF++)~\cite{FF++}, Celeb-DF~\cite{Celeb} and DFDC\cite{DFDC}. For the FF++ dataset, we randomly select $10000$ real face images and $10000$ fake face images from the DeepFake set. Other datasets follow the same operation. We extract the faces from these images and resize them to $512 \times 512$.  

\smallskip
\noindent\textbf{Evaluation Metrics.}
We utilize multiple metrics for evaluation. To evaluate the attack effectiveness, we use Original Accuracy (OA), Attack Success Rate (ASR), and Benign Accuracy (BA). The higher ASR and BA indicate the better attack performance. To measure the visual quality of poisoned faces, we employ SSIM, PSNR, and FID metrics, where higher values of these metrics correspond to better visual quality.

\smallskip
\noindent\textbf{Deepfake Detectors.}
Our method is studied on four mainstream base networks: ResNet50~\cite{ResNet}, EfficientNet-b4~\cite{EfficientNet}, DenseNet~\cite{DenseNet}, and MobileNet~\cite{MobileNet}. These networks are trained on Deepfake datasets with a binary classifier. In addition, we evaluate our method using four state-of-the-art dedicated Deepfake detectors: F3Net~\cite{F3Net}, SRM~\cite{SRM}, NPR~\cite{NPR}, and FG~\cite{FG}

\smallskip
\noindent\textbf{Implementation Details.}
Our method is implemented using PyTorch 2.0.1~\cite{pytorch} with Nvidia 3090 GPUs. During training the trigger generator, we use the batch size of 4 and the Adam optimizer with a learning rate of 1e-4. The weighting factors are set as follows: $\lambda_{dis} = 2, \lambda_{rec} = 1.5, \lambda_{sup} = 1$. The input passcode is a 100-bit binary string. For training the clean-label trigger generator, we employ ResNet as the deepfake detector $\bm{F}$. The poison rate is set to 5\%\footnote{Note that each dataset contains an equal number of real and fake images. We poison 10\% of one class of samples in the main experiment.}

\subsection{Results}

\smallskip
\noindent\textbf{Passcode-controlled Dirty-label Attack.}
Table \ref{tab:pas_dirty} shows the results under this scenario. Note that OA denotes the Original Accuracy without data poisoning. $\bm{p}'_{\alpha}$ indicates a randomly selected passcode in $\mathcal{D}_{aux}$. $\bm{p}'_{\beta}$ indicates a randomly selected passcode not in $\mathcal{D}_{aux}$. $\bm{p}'_{\gamma}$ represents a randomly selected passcode similar to $\bm{p}$, such as $124$ when $\bm{p} = 123$. Specifically, 10\% of real images are poisoned as $\mathcal{D}_{s}$ and assigned with fake labels. 50\% of the real images are set as $\mathcal{D}_{aux}$. {During the training phase, we used 10 different passcodes (excluding $\bm{p}$) in the $\mathcal{D}_{aux}$. During the testing phase, we also used 10 different $\bm{p}'_{\alpha}$, 10 different $\bm{p}'_{\beta}$, and 5 different $\bm{p}'_{\gamma}$ that were very similar to $\bm{p}$.} The results show that the ASR is high when using the correct passcode, but if the passcode is incorrect—even if similar to the correct one—the backdoor cannot be activated, resulting in a significantly low ASR. Also, BA matches with OA for all results, demonstrating the good \textit{Function preservation}. 

Moreover, we validate the generalizability of our method in two configurations: (1) A $\bm{\rightarrow}$ B denotes using the trigger generator trained on A to generate triggers given B and the Deepfake detector is trained and tested on B. (2) A $\bm{\Rightarrow}$ B denotes using the trigger generator trained on A to generate triggers given B and the Deepfake detector is trained on A and tested on B. Table~\ref{tab:ASR_dirty} presents the results for both configurations. Our method achieves high ASR on all methods, indicating good generalizability of our trigger generator. Note that OA and BA drop notably in A $\bm{\Rightarrow}$ B configuration, which is because the Deepfake detector trained on A is limited to identify B.

\smallskip
\noindent\textbf{Passcode-controlled Clean-label Attack.}
In this setting, we poison 10\% of the real images as $\mathcal{D}_{s}$ and set 50\% of the fake images as $\mathcal{D}_{aux}$. Table~\ref{tab:pas_clean} and Table~\ref{tab:ASR_clean} shows the results, revealing a similar trend as in Table~\ref{tab:pas_dirty} and Table~\ref{tab:ASR_dirty}. Although the target label is equal to the true label, the ASR remains high in all scenarios ($\bm{\rightarrow}$ and $\bm{\Rightarrow}$). Meanwhile, BA also aligns with OA in most cases, demonstrating good function preservation as well.

\begin{table*}[!ht]
    \centering
    \small
    \caption{\textbf{Performance (\%) of passcode-controlled clean-label attack.} $\bm{\rightarrow}$ and $\bm{\Rightarrow}$ have the same definition as in Table~\ref{tab:ASR_dirty}.}\vspace{-0.3cm}
    \begin{tabular}{c|ccc|ccc||ccc|ccc} \hline  
    \multirow{2}{*}{Method}  &  \multicolumn{3}{c|}{FF++ $\bm{\rightarrow}$ Celeb-DF} &  \multicolumn{3}{c||}{FF++ $\bm{\rightarrow}$ DFDC} &  \multicolumn{3}{c|}{FF++ $\bm{\Rightarrow}$ Celeb-DF} &  \multicolumn{3}{c}{FF++ $\bm{\Rightarrow}$ DFDC} \\ 
    \cline{2-13}
    &  OA&BA & ASR &  OA&BA & ASR &  OA&BA & ASR &  OA&BA & ASR \\
    \hline  
    ResNet~\cite{ResNet} &  86.67&86.33& 92.35&  75.12&79.58& 96.85&  59.05&65.10& 99.95&  58.02&61.43& 94.50\\ 
    EfficientNet~\cite{EfficientNet} &  89.47&88.48& 97.85&  83.02&82.48& 97.45&  54.18&57.95& 100&  56.08&60.60& 87.85\\  
    MobilNet~\cite{MobileNet} &  86.40&87.86& 98.75&  80.00&80.80& 91.75&  60.15&55.50& 100&  47.90&47.40& 99.90\\  
    DenseNet~\cite{DenseNet} &  85.77&86.90& 98.70&  80.70&80.78& 99.20&  59.65&58.45& 100&  53.28&59.13& 99.60\\ \hline  
    F3Net~\cite{F3Net} &  83.42&89.25& 99.35&  75.25&84.20& 96.05&  61.43&56.45& 100&  62.18&57.20& 98.95\\  
    SRM~\cite{SRM} &  88.55&88.05& 90.25&  81.22&81.20& 94.70&  58.18&55.80& 99.95&  53.65&51.95& 92.65\\  
    NPR~\cite{NPR} &  80.32&79.03& 100&  72.20&72.70& 99.00&  51.18&57.30& 100&  53.90&52.62& 96.10\\   
    FG~\cite{FG} &  90.10&91.83& 98.00&  84.25&86.90& 94.35&  60.43&56.47& 100&  59.08&53.52& 92.50\\ \hline
    \end{tabular}
    \label{tab:ASR_clean}
\end{table*}

\begin{table*}[!ht]
    \centering
    \small
    \caption{Compared with other backdoor attack methods under clean-label scenario.}
    \vspace{-0.3cm}
    \begin{tabular}{c|cc|cc|cc|cc|cc}\hline
    \multirow{2}{*}{Method} & \multicolumn{2}{c}{ResNet}& \multicolumn{2}{|c}{EfficientNet} &  \multicolumn{2}{|c}{F3Net}& \multicolumn{2}{|c}{SRM} & \multicolumn{2}{|c}{Avg.} \\\cline{2-11}
     &  BA&  ASR&  BA&  ASR& BA& ASR& BA&ASR & BA&ASR \\ \hline 
     BadNet~\cite{BadNet} &  98.10&  73.08&  98.79&  83.89 & 95.96& 81.16& 95.30&24.75 & 97.04& 65.72\\ 
     Blended~\cite{Blend} &  97.83&  95.81&  98.59&  \textbf{99.04} & 96.49& 95.30& 95.20&93.78 & 97.03& \underline{95.98}\\  
    SIG~\cite{SIG} & 98.38& 88.28& 98.94& 97.02 & 96.03& 96.16& 95.25&90.40 & 97.15& 92.97\\
    LC~\cite{LC} & 97.57& 87.02& 98.28& 94.34 & 97.35& 95.50& 96.80&77.48 & 97.50& 88.59\\
    \hline
     ISSBA~\cite{ISSBA} &  96.29&  75.15&  98.54&  93.08 & 96.21& 78.49& 96.89&78.23 & 96.98& 81.24\\ 
     PFF~\cite{PFF} & 97.30& 86.82& 98.05& \underline{98.53} & 97.02& 97.52& 97.87&85.30 & 97.56& 92.05\\
     \textbf{Ours} & 97.45& \textbf{96.62}& 98.89& 97.27 & 96.64& \textbf{98.99}& 96.10&\textbf{96.16} & 97.27& \textbf{97.26}\\\hline
    \end{tabular}   
    \label{tab:compare_with_other_attacks}
\end{table*}

\begin{figure*}[!ht]
    \centering
    \includegraphics[width=0.8\linewidth]{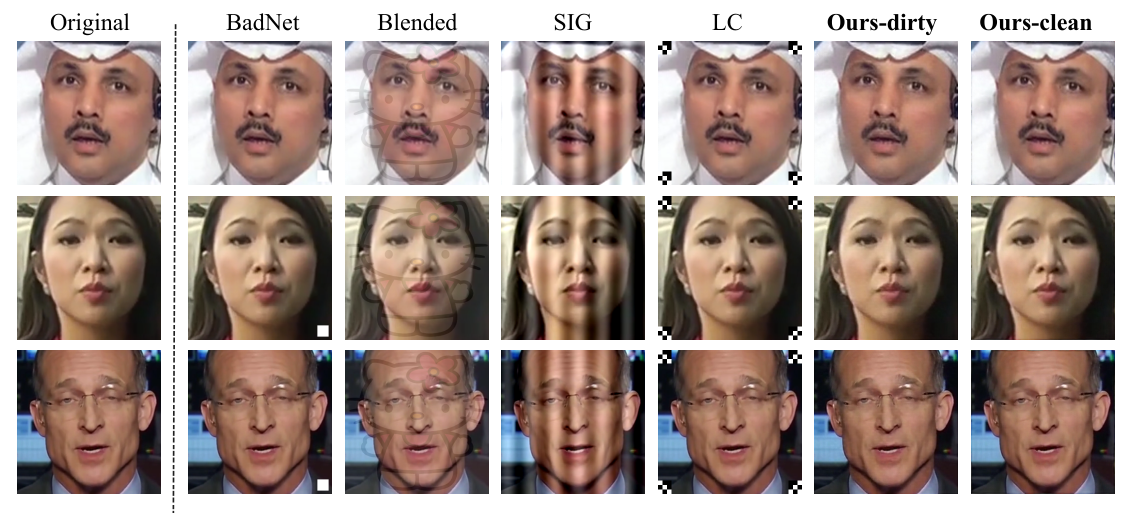}
    \vspace{-0.3cm}
    \caption{Visual comparison with visible trigger methods.}
    \label{fig:visible-trigger}
    \vspace{-0.3cm}
\end{figure*}

\begin{figure}[!ht]
    \centering
    \includegraphics[width=\linewidth]{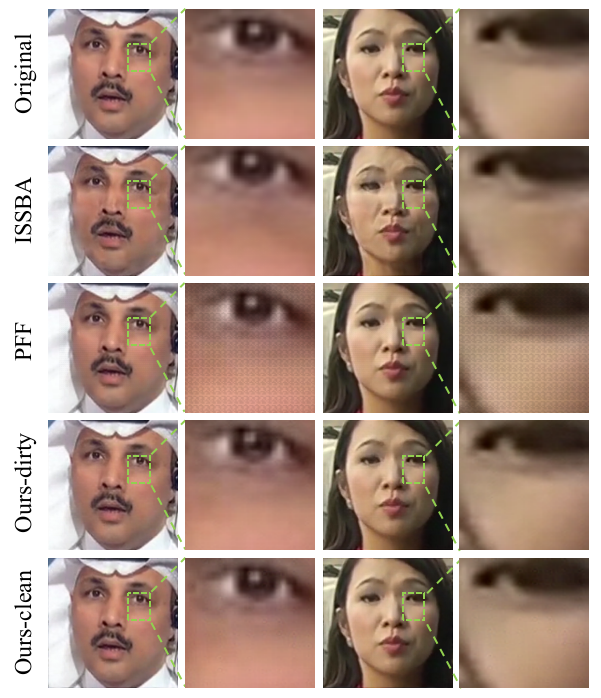}
    \vspace{-0.7cm}
    \caption{Visual comparison with invisible trigger methods.}
    \label{fig:invisible-trigger}
\end{figure}

\smallskip
\noindent\textbf{Compared with Backdoor Attack Methods.}
We compare our method with six backdoor attack methods: BadNet~\cite{BadNet}, Blended~\cite{Blend}, ISSBA~\cite{ISSBA}, SIG~\cite{SIG}, LC~\cite{LC} and PFF~\cite{PFF}. The implementation details are shown in the \textit{Supplementary}. Note that these methods cannot conduct passcode-controlled attacks, so we evaluate them directly under a clean-label attack scenario for a fair comparison. The poison rate is set to 5\% as before.
Table~\ref{tab:compare_with_other_attacks} shows the results of these backdoor attack methods using four Deepfake detectors on FF++ dataset. 
Among them, BadNet~\cite{BadNet}, Blended~\cite{Blend}, SIG~\cite{SIG}, LC~\cite{LC} use visible triggers (see Fig.~\ref{fig:visible-trigger}), while ISSBA~\cite{ISSBA}, PFF~\cite{PFF} and our methods utilize invisible triggers (see Fig.~\ref{fig:invisible-trigger}). Despite this, our method can still achieve the best average performance compared to both visible and invisible trigger methods. 

Fig.~\ref{fig:visible-trigger} and Fig.~\ref{fig:invisible-trigger} provide a visual comparison. Notably, our method surpasses the visual quality of PFF, which exhibits visible mesh grids, and achieves competitive visual quality compared to ISSBA. Table \ref{tab:invisible_score} also evaluates the quality of invisible triggers, showing that our method performs competitively or even surpasses others in most metrics\footnote{These metrics could not fully reflect the true visual quality due to the inner difficulty of quality assessment~\cite{dohmen2024five}, we only use them as reference.}.

\begin{table}
    \centering
    \small
\caption{Quality evaluation for invisible triggers.}
\vspace{-0.3cm}
\label{tab:invisible_score}
    \begin{tabular}{c|c|c|c}
    \hline
         Method&  PSNR$\uparrow$&  SSIM$\uparrow$& FID$\downarrow$\\
         \hline
         ISSBA~\cite{ISSBA}&  30.48&  0.9232& 78.76\\
         PFF~\cite{PFF}&  \underline{33.96}&  0.7807& \textbf{6.86}\\
         \hline
         Ours-dirty&  \textbf{35.71}&  \textbf{0.9572}& \underline{11.78}\\
 Ours-clean& 29.89& \underline{0.9501}&41.65\\
 \hline
    \end{tabular}
    \vspace{-0.3cm}
\end{table}

\subsection{Analysis}

\smallskip
\noindent\textbf{Effect of Deepfake Detector in Clean-label Trigger Generation.} 
As described in Sec.~\ref{sec:trigger-gen}, to generate clean-label triggers, we employ a Deepfake detector $\bm{F}$ (see Fig.~\ref{fig:trigger-train}) to suppress the original representation of input faces.  In this part, we study the effect of using various Deepfake detectors. None denotes not using a Deepfake detector in the training of the trigger generator and directly adding the triggers into samples without altering their true labels. Table~\ref{tab:effect-F} presents the performance of using various Deepfake detectors in the training of the trigger generator. It can be seen that without using the Deepfake detector in training, we can only achieve a decent ASR around $85\%$ on four victim detectors. By using Deepfake detectors in training, the ASR is notably increased. In particular, ResNet-based $\bm{F}$ exhibits great transferability across various victim detectors compared to EfficientNet-based $\bm{F}$. Therefore, we employ ResNet as our $\bm{F}$ in the main experiments.

\begin{table}[!t]
    \centering
    \small
\caption{Effect of Deepfake detector in clean-label trigger generation.}
\vspace{-0.3cm}
\label{tab:detector_loss_comparison}
    \begin{tabular}{c|c|c|c} \hline 
         Victim detector $\downarrow$, $\bm{F} \rightarrow$ &  None &  EfficientNet& ResNet\\ \hline 
         ResNet~\cite{ResNet} &  86.97&  91.91& \textbf{96.62}\\ \hline 
         EfficientNet~\cite{EfficientNet} &  86.92&  93.59& \textbf{97.72}\\ \hline 
         MobileNet~\cite{MobileNet} &  83.64&  90.99& \textbf{96.62}\\ \hline 
         DenseNet~\cite{DenseNet} &  84.14&  91.02& \textbf{96.01}\\ \hline
    \end{tabular}
    \label{tab:effect-F}
\end{table}

\begin{figure}[!t]
    \centering
    \includegraphics[width=1.0\linewidth]{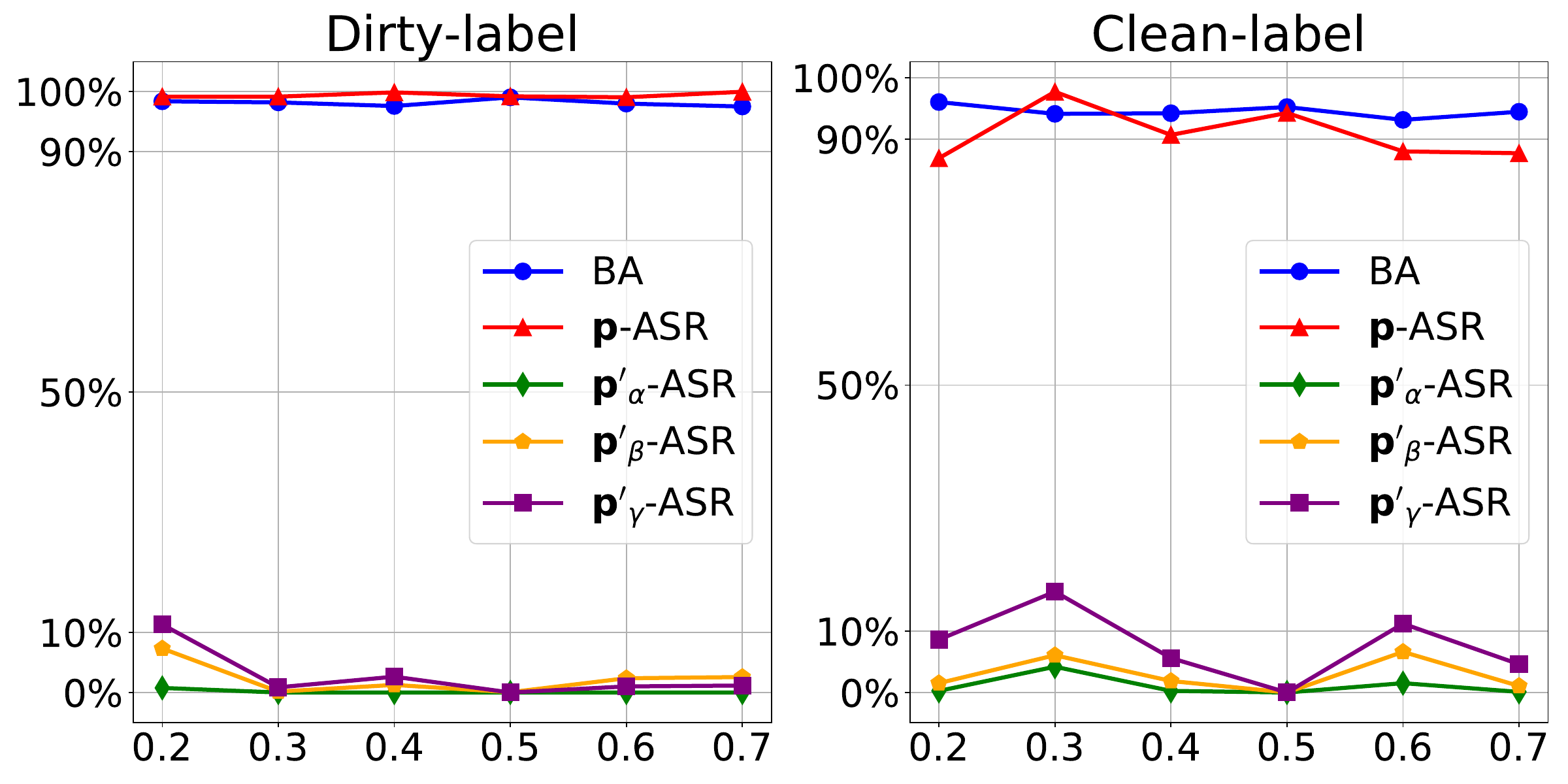}
    \vspace{-0.7cm}
    \caption{Various Scale of $\mathcal{D}_{aux}$.}
    \label{fig:non_passcode_rate}
\end{figure}

\smallskip
\noindent\textbf{Various Scale of $\mathcal{D}_{aux}$.} 
To accomplish the passcode-controlled attack, we require a set of $\mathcal{D}_{aux}$ together with the poisoned set $\mathcal{D}_{p}$. This part studies whether the scale of $\mathcal{D}_{aux}$ affects the performance of our method. For dirty-label attack, we examine various scales ranging from 20\% to 70\% to the real images in $\mathcal{D}_{train} \backslash \mathcal{D}_{s}$. For a clean-label attack, we use the same range but on the fake images in $\mathcal{D}_{train} \backslash \mathcal{D}_{s}$. Fig.~\ref{fig:non_passcode_rate} shows the results under this setting on the ResNet detector. It can be observed that (1) the proportion of different scales of $\mathcal{D}_{aux}$ has a negligible effect on the BA, which means this set would not disturb the essential distribution of the training set. (2) ASR in the dirty-label scenario is almost unchanged. Despite fluctuations in the clean-label scenario, the ASR stays high with various scales. (3) The other passcodes $\bm{p}'_{\alpha}$, $\bm{p}'_{\beta}$, and $\bm{p}'_{\gamma}$ still do not work, resulting in stable low ASR. In main experiments, we use 50\% as the scale of $\mathcal{D}_{aux}$.


\begin{figure}[!t]
    \centering
    \includegraphics[width=\linewidth]{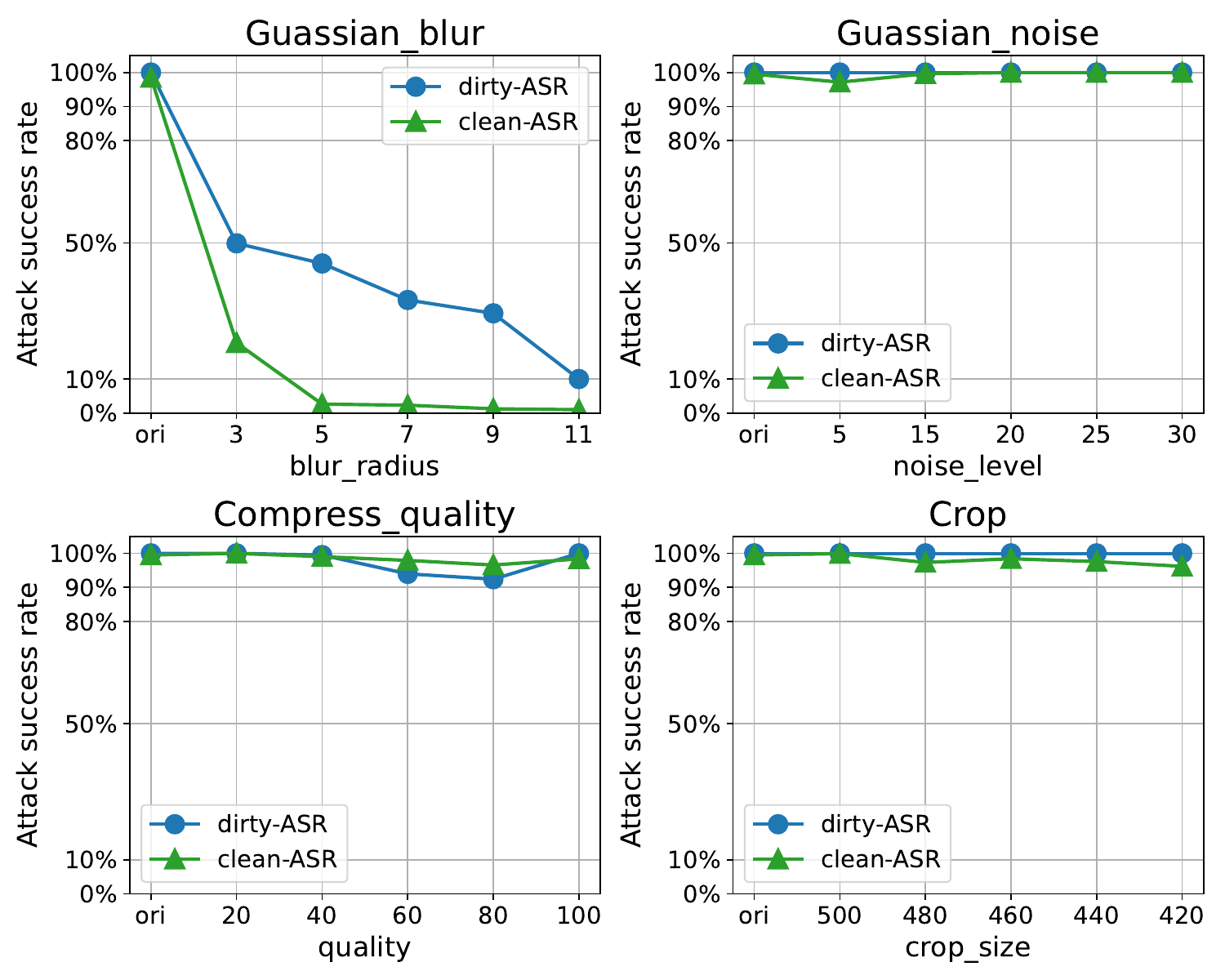}
    \vspace{-0.7cm}
    \caption{ASR under varying degrees of Gaussian blur, Gaussian noise, image compression, and crop operations. }
    \label{fig:trigger_robustness}
\end{figure}

\begin{table}[!t]
    \centering
    \small
    \caption{Performance (\%) of our method against defense methods.}
    \vspace{-0.3cm}
    \label{tab:defense}
    \begin{tabular}{c|c|c|c|c} \hline 
    \multirow{2}{*}{Defense}
 & \multicolumn{2}{c}{clean-label}& \multicolumn{2}{|c}{dirty-label}  \\\cline{2-5} 
         &  ASR&  ACC&  ASR& ACC\\ \hline 
         original&  96.62&  97.45&  100& 98.74\\ 
         FT~\cite{FT}&  98.68&  98.76&  99.19& 98.46\\ 
         FP~\cite{FP}&  88.99&  97.78&  99.84& 97.15\\  
         NAD~\cite{NAD}&  94.49&  97.32&  100& 98.71\\ 
         ABL~\cite{ABL}&  84.19&  98.66&  100& 98.33\\ \hline
    \end{tabular}
\end{table}

\begin{figure*}[!t]
    \centering
    \includegraphics[width=0.95\textwidth]{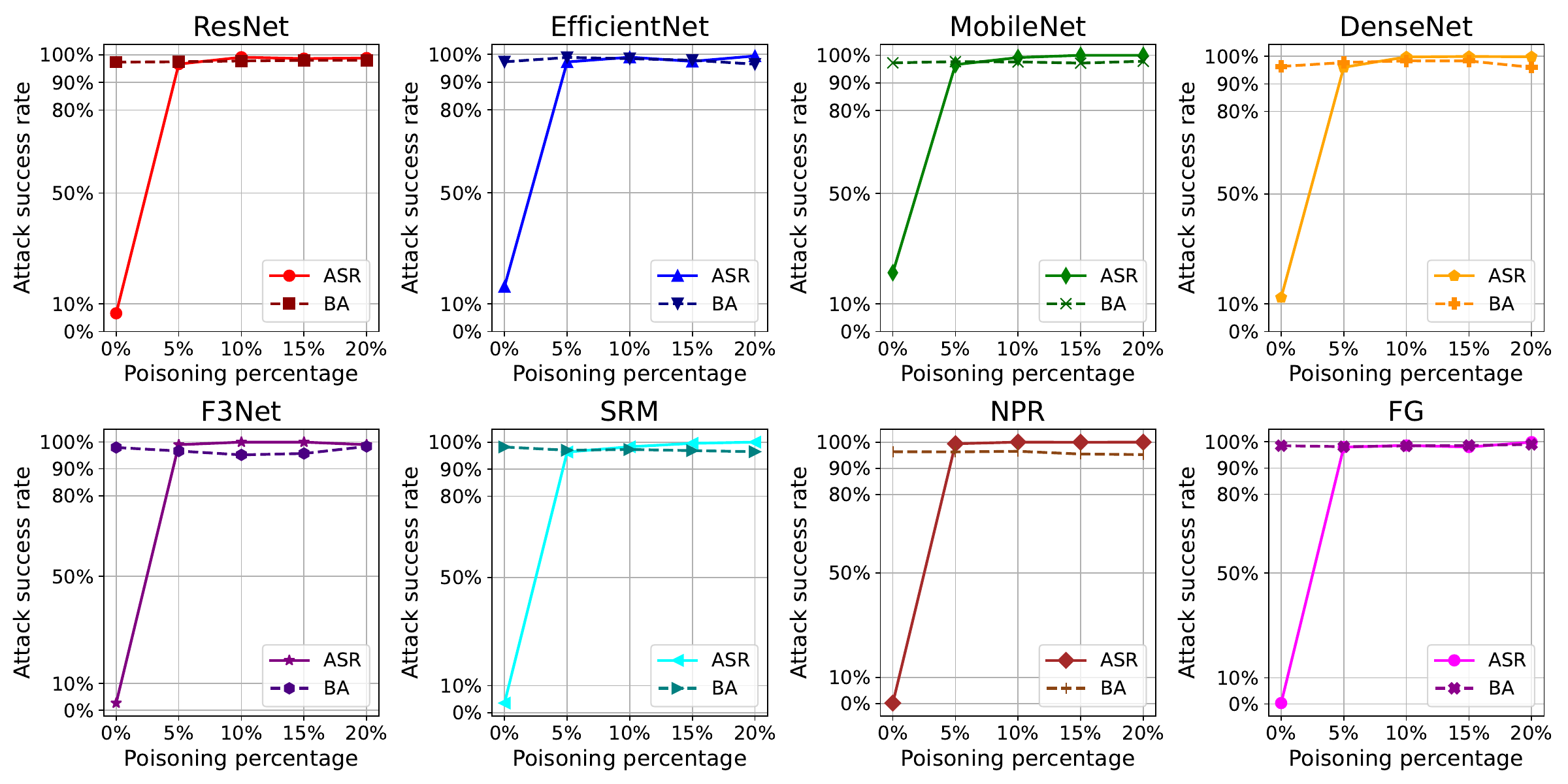}
    \vspace{-0.4cm}
    \caption{Performance under different poisoning ratios in the clean-label scenario.}
    \label{fig:poison_rate_ASR}
    \vspace{-0.3cm}
\end{figure*}

\smallskip
\noindent\textbf{Trigger Robustness.} 
This part verifies the robustness of the proposed trigger on detector F3Net. Specifically, we apply four image distortions of Gaussian blur, Gaussian noise, image compression, and cropping, under both dirty-label and clean-label scenarios. For Gaussian blur, we vary the blur radius within the range $[3,5,7,9,11]$. For Gaussian noise, we set the mean to 0 and the standard deviation range to $[5,15,20,25,30]$. The image compression factors are in $[20,40,60,80,100]$. Given the original image size of $512$, we crop face images at different sizes, ranging from $[500,480,460,440,420]$. The results are shown in Fig.~\ref{fig:trigger_robustness}. The findings show that the generated trigger is particularly sensitive to Gaussian blur, with the ASR dropping rapidly as the blur radius increases. At a radius of 11, the trigger's effect is almost entirely negated. Nevertheless, the trigger demonstrates strong resilience to other operations. We hypothesize that this is because the averaging effect in blur operations significantly disrupts the trigger pattern, while operations like adding noise and cropping primarily affect local content, having minimal impact on the overall structure of the trigger. Interestingly, image compression has a minimal effect, which is likely because compression merely distorts the trigger pattern.

\smallskip
\noindent\textbf{Resistance to Defenses.} 
In this part, we evaluate the resistance of our method against various backdoor
defenses, including FT~\cite{FT}, FP~\cite{FP}, NAD~\cite{NAD} and ABL~\cite{ABL}. For the backdoor defense setup, we follow the protocols used in~\cite{backdoorbench}. Specifically, for FT, we fine-tune the victim detector using 5\% clean data. For FP, we prune 99\% of the neurons in the last convolutional layer of the victim detector and subsequently fine-tune it using 5\% clean data. For NAD, we use the victim detector fine-tuned on 5\% clean data as the teacher model and implement distillation on the original victim detector. For ABL, we isolate 10\% of suspicious data and conduct the backdoor unlearning using the default setting. As shown in Table~\ref{tab:defense}, these defense methods fail to effectively counteract our method, as ASR scores remain high in both dirty-label and clean-label scenarios. This outcome is likely due to (1) the defense methods being tailored to general classification tasks, making them unsuitable for Deepfake detection, and (2) the unique properties of our trigger, which can evade these defenses. These findings highlight the potential of our method to evade the Deepfake detectors in practice.

\smallskip
\noindent\textbf{Various Poisoning Rates.} 
Since the dirty-label attacks can easily achieve 100\% ASR by only poisoning 5\% of the training set, we mainly discuss the effect of using various poison rates in the clean-label attack scenario. Specifically, we experiment with poisoning rates of 0\%, 5\%, 10\%, 15\%, and 20\%. Fig.~\ref{fig:poison_rate_ASR} presents the results, showing that as the poisoning rate increases, the BA score remains relatively stable, suggesting that the Deepfake detectors are robust and well-trained. By only using a poisoning rate of 5\%, the ASR score quickly rises to the maximum and becomes smooth as the rate increases further, demonstrating the effectiveness of our method even with minimal poisoning.

\smallskip
\noindent\textbf{Limitations.} 
As studied in the robustness section, our method is relatively fragile to the blurring operations. Nevertheless, heavy blurring can also notably degrade the visual quality of faces, meaning such operations are unlikely to be commonly used in practice.

\section{Conclusion}


This paper investigates a new security risk that can seriously undermine the trustfulness of Deepfake detectors and describes solutions to subtly infect Deepfake detectors. Specifically, we develop a trigger generator capable of creating passcode-controlled, representation-suppression, adaptive, and invisible trigger patterns. Then we discuss two poisoning scenarios, dirty-label poisoning and clean-label poisoning, to add these triggers into training images, which can effectively inject backdoors into Deepfake detectors. Extensive experiments highlight the effectiveness, stealthiness, and practicality of our method compared to several baselines. We aim to raise awareness within the forensics community about the security risks posed by third-party data providers.

\smallskip
\noindent\textbf{Acknowledgement.} This work is supported in part by the National Natural Science Foundation of China (No.62402464), Shandong Provincial Natural Science Foundation (No.ZR2024QF035), and China Postdoctoral Science Foundation (No.2021TQ0314; No.2021M703036).


{
    \small
    \bibliographystyle{ieeenat_fullname}
    \bibliography{main}

\begin{thebibliography}{51}
\providecommand{\natexlab}[1]{#1}
\providecommand{\url}[1]{\texttt{#1}}
\expandafter\ifx\csname urlstyle\endcsname\relax
  \providecommand{\doi}[1]{doi: #1}\else
  \providecommand{\doi}{doi: \begingroup \urlstyle{rm}\Url}\fi

\bibitem[Al-Khazraji et~al.(2023)Al-Khazraji, Saleh, KHALID, and MISHKHAL]{social_concern02}
Samer~Hussain Al-Khazraji, Hassan~Hadi Saleh, Adil~Ibrahim KHALID, and Israa~Adnan MISHKHAL.
\newblock Impact of deepfake technology on social media: Detection, misinformation and societal implications.
\newblock \emph{The Eurasia Proceedings of Science Technology Engineering and Mathematics}, 23:\penalty0 429--441, 2023.

\bibitem[Barni et~al.(2019)Barni, Kallas, and Tondi]{SIG}
Mauro Barni, Kassem Kallas, and Benedetta Tondi.
\newblock A new backdoor attack in cnns by training set corruption without label poisoning.
\newblock In \emph{2019 IEEE International Conference on Image Processing (ICIP)}, pages 101--105. IEEE, 2019.

\bibitem[Carlini and Farid(2020)]{evading_deepfake_detectors03}
Nicholas Carlini and Hany Farid.
\newblock Evading deepfake-image detectors with white-and black-box attacks.
\newblock In \emph{Proceedings of the IEEE/CVF conference on computer vision and pattern recognition workshops}, pages 658--659, 2020.

\bibitem[Chen et~al.(2017)Chen, Liu, Li, Lu, and Song]{Blend}
Xinyun Chen, Chang Liu, Bo Li, Kimberly Lu, and Dawn Song.
\newblock Targeted backdoor attacks on deep learning systems using data poisoning.
\newblock \emph{arXiv preprint arXiv:1712.05526}, 2017.

\bibitem[Chesney and Citron(2019)]{political_misinfor}
Robert Chesney and Danielle Citron.
\newblock Deepfakes and the new disinformation war: The coming age of post-truth geopolitics.
\newblock \emph{Foreign Aff.}, 98:\penalty0 147, 2019.

\bibitem[Dohmen et~al.(2024)Dohmen, Truong, Baltruschat, and Lenga]{dohmen2024five}
Melanie Dohmen, Tuan Truong, Ivo~M Baltruschat, and Matthias Lenga.
\newblock Five pitfalls when assessing synthetic medical images with reference metrics.
\newblock In \emph{MICCAI Workshop on Deep Generative Models}, pages 150--159. Springer, 2024.

\bibitem[Dolhansky et~al.(2020)Dolhansky, Bitton, Pflaum, Lu, Howes, Wang, and Ferrer]{DFDC}
Brian Dolhansky, Joanna Bitton, Ben Pflaum, Jikuo Lu, Russ Howes, Menglin Wang, and Cristian~Canton Ferrer.
\newblock The deepfake detection challenge (dfdc) dataset.
\newblock \emph{arXiv preprint arXiv:2006.07397}, 2020.

\bibitem[Frank et~al.(2020)Frank, Eisenhofer, Sch{\"o}nherr, Fischer, Kolossa, and Holz]{frank2020leveraging}
Joel Frank, Thorsten Eisenhofer, Lea Sch{\"o}nherr, Asja Fischer, Dorothea Kolossa, and Thorsten Holz.
\newblock Leveraging frequency analysis for deep fake image recognition.
\newblock In \emph{International conference on machine learning}, pages 3247--3258. PMLR, 2020.

\bibitem[Gamage et~al.(2022)Gamage, Ghasiya, Bonagiri, Whiting, and Sasahara]{social_concern01}
Dilrukshi Gamage, Piyush Ghasiya, Vamshi Bonagiri, Mark~E Whiting, and Kazutoshi Sasahara.
\newblock Are deepfakes concerning? analyzing conversations of deepfakes on reddit and exploring societal implications.
\newblock In \emph{Proceedings of the 2022 CHI Conference on Human Factors in Computing Systems}, pages 1--19, 2022.

\bibitem[Golda et~al.(2024)Golda, Mekonen, Pandey, Singh, Hassija, Chamola, and Sikdar]{privacy_invasion}
Abenezer Golda, Kidus Mekonen, Amit Pandey, Anushka Singh, Vikas Hassija, Vinay Chamola, and Biplab Sikdar.
\newblock Privacy and security concerns in generative ai: A comprehensive survey.
\newblock \emph{IEEE Access}, 2024.

\bibitem[Goodfellow et~al.(2014)Goodfellow, Shlens, and Szegedy]{goodfellow2014explaining}
Ian~J Goodfellow, Jonathon Shlens, and Christian Szegedy.
\newblock Explaining and harnessing adversarial examples.
\newblock \emph{arXiv preprint arXiv:1412.6572}, 2014.

\bibitem[Gu et~al.(2017)Gu, Dolan-Gavitt, and Garg]{BadNet}
Tianyu Gu, Brendan Dolan-Gavitt, and Siddharth Garg.
\newblock Badnets: Identifying vulnerabilities in the machine learning model supply chain.
\newblock \emph{arXiv preprint arXiv:1708.06733}, 2017.

\bibitem[Guo et~al.(2023)Guo, Zhen, and Yan]{guo2023controllable}
Ying Guo, Cheng Zhen, and Pengfei Yan.
\newblock Controllable guide-space for generalizable face forgery detection.
\newblock In \emph{Proceedings of the IEEE/CVF International Conference on Computer Vision}, pages 20818--20827, 2023.

\bibitem[Hancock and Bailenson(2021)]{social_concern03}
Jeffrey~T Hancock and Jeremy~N Bailenson.
\newblock The social impact of deepfakes, 2021.

\bibitem[He et~al.(2016)He, Zhang, Ren, and Sun]{ResNet}
Kaiming He, Xiangyu Zhang, Shaoqing Ren, and Jian Sun.
\newblock Deep residual learning for image recognition.
\newblock In \emph{Proceedings of the IEEE conference on computer vision and pattern recognition}, pages 770--778, 2016.

\bibitem[Hou et~al.(2023)Hou, Guo, Huang, Xie, Ma, and Zhao]{evading_deepfake_detectors01}
Yang Hou, Qing Guo, Yihao Huang, Xiaofei Xie, Lei Ma, and Jianjun Zhao.
\newblock Evading deepfake detectors via adversarial statistical consistency.
\newblock In \emph{Proceedings of the IEEE/CVF Conference on Computer Vision and Pattern Recognition}, pages 12271--12280, 2023.

\bibitem[Howard(2017)]{MobileNet}
Andrew~G Howard.
\newblock Mobilenets: Efficient convolutional neural networks for mobile vision applications.
\newblock \emph{arXiv preprint arXiv:1704.04861}, 2017.

\bibitem[Huang et~al.(2017)Huang, Liu, Van Der~Maaten, and Weinberger]{DenseNet}
Gao Huang, Zhuang Liu, Laurens Van Der~Maaten, and Kilian~Q Weinberger.
\newblock Densely connected convolutional networks.
\newblock In \emph{Proceedings of the IEEE conference on computer vision and pattern recognition}, pages 4700--4708, 2017.

\bibitem[Huang et~al.(2020)Huang, Juefei-Xu, Guo, Xie, Ma, Miao, Liu, and Pu]{evading_deepfake_detectors02}
Yihao Huang, Felix Juefei-Xu, Qing Guo, Xiaofei Xie, Lei Ma, Weikai Miao, Yang Liu, and Geguang Pu.
\newblock Fakeretouch: Evading deepfakes detection via the guidance of deliberate noise.
\newblock \emph{arXiv preprint arXiv:2009.09213}, 1\penalty0 (2), 2020.

\bibitem[Jung et~al.(2020)Jung, Kim, and Kim]{biological_signals02}
Tackhyun Jung, Sangwon Kim, and Keecheon Kim.
\newblock Deepvision: Deepfakes detection using human eye blinking pattern.
\newblock \emph{IEEE Access}, 8:\penalty0 83144--83154, 2020.

\bibitem[Khan et~al.(2024)Khan, Taqi, and Afzal]{economic_fraud}
Rizwan Khan, Mohd Taqi, and Atif Afzal.
\newblock Deepfakes in finance: Unraveling the threat landscape and detection challenges.
\newblock In \emph{Navigating the World of Deepfake Technology}, pages 91--120. IGI Global, 2024.

\bibitem[Kirkpatrick et~al.(2017)Kirkpatrick, Pascanu, Rabinowitz, Veness, Desjardins, Rusu, Milan, Quan, Ramalho, Grabska-Barwinska, et~al.]{FT}
James Kirkpatrick, Razvan Pascanu, Neil Rabinowitz, Joel Veness, Guillaume Desjardins, Andrei~A Rusu, Kieran Milan, John Quan, Tiago Ramalho, Agnieszka Grabska-Barwinska, et~al.
\newblock Overcoming catastrophic forgetting in neural networks.
\newblock \emph{Proceedings of the national academy of sciences}, 114\penalty0 (13):\penalty0 3521--3526, 2017.

\bibitem[Li et~al.(2020{\natexlab{a}})Li, Bao, Zhang, Yang, Chen, Wen, and Guo]{li2020face}
Lingzhi Li, Jianmin Bao, Ting Zhang, Hao Yang, Dong Chen, Fang Wen, and Baining Guo.
\newblock Face x-ray for more general face forgery detection.
\newblock In \emph{Proceedings of the IEEE/CVF conference on computer vision and pattern recognition (CVPR)}, pages 5001--5010, 2020{\natexlab{a}}.

\bibitem[Li and Lyu(2019)]{Li_2019_CVPR_Workshops}
Yuezun Li and Siwei Lyu.
\newblock Exposing deepfake videos by detecting face warping artifacts.
\newblock In \emph{Proceedings of the IEEE/CVF Conference on Computer Vision and Pattern Recognition (CVPR) Workshops}, 2019.

\bibitem[Li et~al.(2020{\natexlab{b}})Li, Yang, Sun, Qi, and Lyu]{Celeb}
Yuezun Li, Xin Yang, Pu Sun, Honggang Qi, and Siwei Lyu.
\newblock Celeb-df: A large-scale challenging dataset for deepfake forensics.
\newblock In \emph{Proceedings of the IEEE/CVF conference on computer vision and pattern recognition}, pages 3207--3216, 2020{\natexlab{b}}.

\bibitem[Li et~al.(2021{\natexlab{a}})Li, Li, Wu, Li, He, and Lyu]{ISSBA}
Yuezun Li, Yiming Li, Baoyuan Wu, Longkang Li, Ran He, and Siwei Lyu.
\newblock Invisible backdoor attack with sample-specific triggers.
\newblock In \emph{Proceedings of the IEEE/CVF international conference on computer vision}, pages 16463--16472, 2021{\natexlab{a}}.

\bibitem[Li et~al.(2021{\natexlab{b}})Li, Lyu, Koren, Lyu, Li, and Ma]{ABL}
Yige Li, Xixiang Lyu, Nodens Koren, Lingjuan Lyu, Bo Li, and Xingjun Ma.
\newblock Anti-backdoor learning: Training clean models on poisoned data.
\newblock \emph{Advances in Neural Information Processing Systems}, 34:\penalty0 14900--14912, 2021{\natexlab{b}}.

\bibitem[Li et~al.(2021{\natexlab{c}})Li, Lyu, Koren, Lyu, Li, and Ma]{NAD}
Yige Li, Xixiang Lyu, Nodens Koren, Lingjuan Lyu, Bo Li, and Xingjun Ma.
\newblock Neural attention distillation: Erasing backdoor triggers from deep neural networks.
\newblock \emph{arXiv preprint arXiv:2101.05930}, 2021{\natexlab{c}}.

\bibitem[Liang et~al.(2024)Liang, Liang, Liu, Jia, Kuang, and Cao]{PFF}
Jiawei Liang, Siyuan Liang, Aishan Liu, Xiaojun Jia, Junhao Kuang, and Xiaochun Cao.
\newblock Poisoned forgery face: Towards backdoor attacks on face forgery detection.
\newblock \emph{arXiv preprint arXiv:2402.11473}, 2024.

\bibitem[Lin et~al.(2024)Lin, He, Ju, Wang, Ding, and Hu]{FG}
Li Lin, Xinan He, Yan Ju, Xin Wang, Feng Ding, and Shu Hu.
\newblock Preserving fairness generalization in deepfake detection.
\newblock In \emph{Proceedings of the IEEE/CVF Conference on Computer Vision and Pattern Recognition}, pages 16815--16825, 2024.

\bibitem[Liu et~al.(2024)Liu, Tan, Tan, Wei, Wang, and Zhao]{liu2024forgery}
Huan Liu, Zichang Tan, Chuangchuang Tan, Yunchao Wei, Jingdong Wang, and Yao Zhao.
\newblock Forgery-aware adaptive transformer for generalizable synthetic image detection.
\newblock In \emph{Proceedings of the IEEE/CVF Conference on Computer Vision and Pattern Recognition}, pages 10770--10780, 2024.

\bibitem[Liu et~al.(2018)Liu, Dolan-Gavitt, and Garg]{FP}
Kang Liu, Brendan Dolan-Gavitt, and Siddharth Garg.
\newblock Fine-pruning: Defending against backdooring attacks on deep neural networks.
\newblock In \emph{International symposium on research in attacks, intrusions, and defenses}, pages 273--294. Springer, 2018.

\bibitem[Luo et~al.(2021)Luo, Zhang, Yan, and Liu]{SRM}
Yuchen Luo, Yong Zhang, Junchi Yan, and Wei Liu.
\newblock Generalizing face forgery detection with high-frequency features.
\newblock In \emph{Proceedings of the IEEE/CVF conference on computer vision and pattern recognition}, pages 16317--16326, 2021.

\bibitem[Paszke et~al.(2019)Paszke, Gross, Massa, Lerer, Bradbury, Chanan, Killeen, Lin, Gimelshein, Antiga, et~al.]{pytorch}
Adam Paszke, Sam Gross, Francisco Massa, Adam Lerer, James Bradbury, Gregory Chanan, Trevor Killeen, Zeming Lin, Natalia Gimelshein, Luca Antiga, et~al.
\newblock Pytorch: An imperative style, high-performance deep learning library.
\newblock \emph{Advances in neural information processing systems}, 2019.

\bibitem[Qi et~al.(2020)Qi, Guo, Juefei-Xu, Xie, Ma, Feng, Liu, and Zhao]{qi2020deeprhythm}
Hua Qi, Qing Guo, Felix Juefei-Xu, Xiaofei Xie, Lei Ma, Wei Feng, Yang Liu, and Jianjun Zhao.
\newblock Deeprhythm: Exposing deepfakes with attentional visual heartbeat rhythms.
\newblock In \emph{Proceedings of the 28th ACM international conference on multimedia}, pages 4318--4327, 2020.

\bibitem[Qian et~al.(2020{\natexlab{a}})Qian, Yin, Sheng, Chen, and Shao]{F3Net}
Yuyang Qian, Guojun Yin, Lu Sheng, Zixuan Chen, and Jing Shao.
\newblock Thinking in frequency: Face forgery detection by mining frequency-aware clues.
\newblock In \emph{European conference on computer vision}, pages 86--103. Springer, 2020{\natexlab{a}}.

\bibitem[Qian et~al.(2020{\natexlab{b}})Qian, Yin, Sheng, Chen, and Shao]{qian2020thinking}
Yuyang Qian, Guojun Yin, Lu Sheng, Zixuan Chen, and Jing Shao.
\newblock Thinking in frequency: Face forgery detection by mining frequency-aware clues.
\newblock In \emph{European conference on computer vision}, pages 86--103. Springer, 2020{\natexlab{b}}.

\bibitem[Ronneberger et~al.(2015)Ronneberger, Fischer, and Brox]{U-Net}
Olaf Ronneberger, Philipp Fischer, and Thomas Brox.
\newblock U-net: Convolutional networks for biomedical image segmentation.
\newblock In \emph{Medical image computing and computer-assisted intervention--MICCAI 2015: 18th international conference, Munich, Germany, October 5-9, 2015, proceedings, part III 18}, pages 234--241. Springer, 2015.

\bibitem[Rossler et~al.(2019)Rossler, Cozzolino, Verdoliva, Riess, Thies, and Nie{\ss}ner]{FF++}
Andreas Rossler, Davide Cozzolino, Luisa Verdoliva, Christian Riess, Justus Thies, and Matthias Nie{\ss}ner.
\newblock Faceforensics++: Learning to detect manipulated facial images.
\newblock In \emph{Proceedings of the IEEE/CVF international conference on computer vision}, pages 1--11, 2019.

\bibitem[Shiohara and Yamasaki(2022)]{shiohara2022detecting}
Kaede Shiohara and Toshihiko Yamasaki.
\newblock Detecting deepfakes with self-blended images.
\newblock In \emph{Proceedings of the IEEE/CVF Conference on Computer Vision and Pattern Recognition (CVPR)}, pages 18720--18729, 2022.

\bibitem[Szegedy(2013)]{szegedy2013intriguing}
C Szegedy.
\newblock Intriguing properties of neural networks.
\newblock \emph{arXiv preprint arXiv:1312.6199}, 2013.

\bibitem[Tan et~al.(2024)Tan, Zhao, Wei, Gu, Liu, and Wei]{NPR}
Chuangchuang Tan, Yao Zhao, Shikui Wei, Guanghua Gu, Ping Liu, and Yunchao Wei.
\newblock Rethinking the up-sampling operations in cnn-based generative network for generalizable deepfake detection.
\newblock In \emph{Proceedings of the IEEE/CVF Conference on Computer Vision and Pattern Recognition}, pages 28130--28139, 2024.

\bibitem[Tan and Le(2019)]{EfficientNet}
Mingxing Tan and Quoc Le.
\newblock Efficientnet: Rethinking model scaling for convolutional neural networks.
\newblock In \emph{International conference on machine learning}, pages 6105--6114. PMLR, 2019.

\bibitem[Tancik et~al.(2020)Tancik, Mildenhall, and Ng]{Stegastamp}
Matthew Tancik, Ben Mildenhall, and Ren Ng.
\newblock Stegastamp: Invisible hyperlinks in physical photographs.
\newblock In \emph{Proceedings of the IEEE/CVF conference on computer vision and pattern recognition}, pages 2117--2126, 2020.

\bibitem[Turner et~al.(2019)Turner, Tsipras, and Madry]{LC}
Alexander Turner, Dimitris Tsipras, and Aleksander Madry.
\newblock Label-consistent backdoor attacks.
\newblock \emph{arXiv preprint arXiv:1912.02771}, 2019.

\bibitem[Wang et~al.(2024)Wang, Liao, Chow, Lin, and Wang]{DeepfakeSurvey2024Wang}
Tianyi Wang, Xin Liao, Kam~Pui Chow, Xiaodong Lin, and Yinglong Wang.
\newblock Deepfake detection: A comprehensive survey from the reliability perspective.
\newblock \emph{ACM Comput. Surv.}, 57\penalty0 (3), 2024.

\bibitem[Wu et~al.(2022)Wu, Chen, Zhang, Zhu, Wei, Yuan, and Shen]{backdoorbench}
Baoyuan Wu, Hongrui Chen, Mingda Zhang, Zihao Zhu, Shaokui Wei, Danni Yuan, and Chao Shen.
\newblock Backdoorbench: A comprehensive benchmark of backdoor learning.
\newblock \emph{Advances in Neural Information Processing Systems}, 35:\penalty0 10546--10559, 2022.

\bibitem[Yan et~al.(2024)Yan, Luo, Lyu, Liu, and Wu]{yan2024transcending}
Zhiyuan Yan, Yuhao Luo, Siwei Lyu, Qingshan Liu, and Baoyuan Wu.
\newblock Transcending forgery specificity with latent space augmentation for generalizable deepfake detection.
\newblock In \emph{Proceedings of the IEEE/CVF Conference on Computer Vision and Pattern Recognition}, pages 8984--8994, 2024.

\bibitem[Yang et~al.(2019)Yang, Li, and Lyu]{yang2019exposing}
Xin Yang, Yuezun Li, and Siwei Lyu.
\newblock Exposing deep fakes using inconsistent head poses.
\newblock In \emph{ICASSP 2019-2019 IEEE International Conference on Acoustics, Speech and Signal Processing (ICASSP)}, pages 8261--8265. IEEE, 2019.

\bibitem[Zhang et~al.(2018)Zhang, Isola, Efros, Shechtman, and Wang]{perceptual}
Richard Zhang, Phillip Isola, Alexei~A Efros, Eli Shechtman, and Oliver Wang.
\newblock The unreasonable effectiveness of deep features as a perceptual metric.
\newblock In \emph{Proceedings of the IEEE conference on computer vision and pattern recognition}, pages 586--595, 2018.

\bibitem[Zhao et~al.(2021)Zhao, Xu, Xu, Ding, Xiong, and Xia]{zhao2021learning}
Tianchen Zhao, Xiang Xu, Mingze Xu, Hui Ding, Yuanjun Xiong, and Wei Xia.
\newblock Learning self-consistency for deepfake detection.
\newblock In \emph{Proceedings of the IEEE/CVF international conference on computer vision (ICCV)}, pages 15023--15033, 2021.

\end{thebibliography}
}


\end{document}